\documentclass[%
 reprint,
 amsmath,amssymb,
 aps,
 reprint,
]{revtex4-1}
\usepackage[most]{tcolorbox}
\usepackage{tikz}
\usepackage{graphicx}% Include figure files
\usepackage{dcolumn}% Align table columns on decimal point
\usepackage{bm}% bold math
\usepackage{relsize}
\usepackage[normalem]{ulem}
\begin{document}

\preprint{AIP/123-QED}

 \title{Unveiling the Spatiotemporal Evolution of Liquid-Lens Coalescence: Self-Similarity, Vortex Quadrupoles, and Turbulence in a Three-Phase Fluid System}% Force line breaks with \\
% \title{The Spatiotemporal Evolution of Liquid-Lens Coalescence: Self-Similarity and Turbulence in a Three-Phase Fluid System}

\author{Nadia Bihari Padhan}
 \email{nadia@iisc.ac.in}
\author{Rahul Pandit}%
 \email{rahul@iisc.ac.in}
\affiliation{%
 Centre for Condensed Matter Theory, Department of Physics, Indian Institute of Science, Bangalore 560012, India. %\textbackslash\textbackslash
}%

\date{\today}% It is always \today, today,

\begin{abstract}
The coalescence of liquid lenses is an important problem at the intersection of fluid dynamics and statistical physics, particularly in the context of complex multi-phase flows. We demonstrate that the three-phase Cahn-Hilliard-Navier-Stokes (CHNS3) system provides a natural theoretical framework for studying liquid-lens coalescence, which has been investigated in recent experiments. Our extensive direct numerical simulations (DNSs) of lens coalescence, in the two and three dimensional (2D and 3D) CHNS3, uncover the rich spatiotemporal evolution of the fluid velocity $\bm u$ and vorticity $\bm \omega$, the concentration fields $c_1, \, c_2,$ and $c_3$ of the three liquids, and an excess pressure $P^G_\mathcal{L}$, which we define in terms of these concentrations via a Poisson equation. We find, in agreement with experiments, that as the lenses coalesce, their neck height $h(t) \sim t^{\alpha_v}$, with $\alpha_v \simeq 1$ in the viscous regime, and $h(t) \sim t^{\alpha_i}$, with $\alpha_i \simeq 2/3$ in the inertial regime. We obtain the crossover from the viscous to the inertial regimes as a function of the  Ohnesorge number $Oh$, a dimensionless combination of viscous stresses and inertial and surface tension forces. We show that a vortex quadrupole, which  straddles the neck of the merging lenses, and $P^G_\mathcal{L}$ play crucial roles in distinguishing between the viscous- and inertial-regime growths of the merging lenses. In the inertial regime we find signatures of turbulence, which we quantify via kinetic-energy and concentration spectra. Finally, we examine the merger of asymmetric lenses, in which the initial stages of coalescence occur along the circular parts of the lens interfaces; in this case, we obtain power-law forms for the $h(t)$ with inertial-regime exponents that lie between their droplet-coalescence and lens-merger counterparts.  

\end{abstract}

\maketitle
\newpage
%%%%%%%%%%%%%%%%%%%%%%%%%%%%%%%%%%%%%%%%%%%%%%%%%%%%%%%%%%%%%%%%%%%

\section{Introduction} 
\begin{figure}
{
\includegraphics[width=8.5cm]
{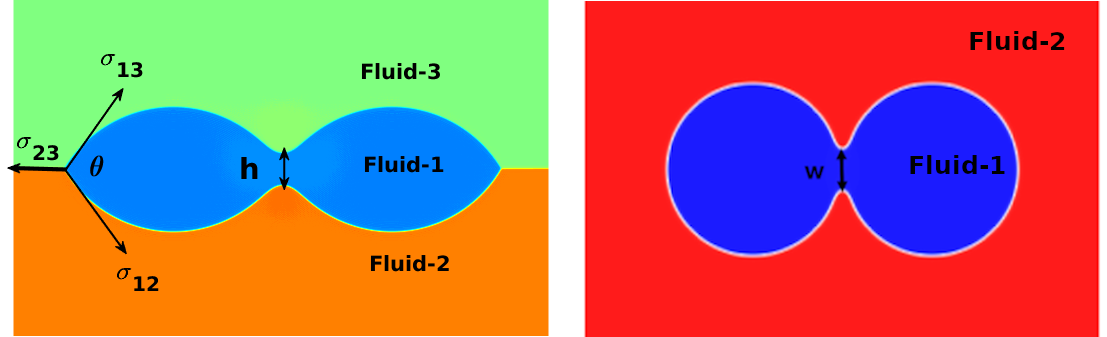}
\put(-240, 65){\rm {\bf(a)}}
\put(-115, 65){\rm {\bf(b)}}
}
\caption{\label{fig:lens} Schematic diagrams illustrating liquid-lens coalescence: (a) 2D or in 3D (a planar section containing the principal axes of the coalescing lenticular biconvex lenses)  and (b) top view in 3D (a planar section perpendicular to the principal axes of the coalescing lenticular biconvex lenses).}
\end{figure}

Coalescence -- of droplets, in general, and liquid lenses, in particular -- is a fundamental problem in the fluid dynamics and statistical physics of multi-phase flows~\cite{fardin2022spreading,Thomas_2004, Dirk_2005, Burton_2007, Paulsen_2011,paulsen2014coalescence,Eggers_1999, Eggers_2003, Volker_2018, Gross_2013, Akella_2020,zhang2023effect,matsuo2023sequentially,li2023dynamics,qian2023inner,heinen2022droplet,huang2022lattice,varma2022elasticity,zhang2022hydrodynamics,xu2022bridge,liu2022laplace,xing2022simulation}. Such droplet merging is of direct relevance in engineering applications, such as ink-jet printers~\cite{hong2023spread,zawadzki20223d}, and atmospheric physics, e.g., the merger of rain drops in a cloud~\cite{pruppacher1978growth,zhang2021experimental,bartlett1966growth}. When two droplets coalesce, a bridge forms and its height $h$  grows with the time $t$. Experiments~\cite{Thomas_2004, Dirk_2005, Burton_2007, Paulsen_2011,paulsen2014coalescence}, theory, and numerical simulations~\cite{Eggers_1999, Eggers_2003, Volker_2018, Gross_2013, Akella_2020} show that, in the early stage of coalescence of two, initially static, spherical droplets, there is self-similar growth with $h(t) \sim t$ and $h(t) \sim t^{1/2}$ in the viscous and inertial regimes, respectively~\cite{Burton_2007, Paulsen_2011, Zhang_2019}. Three-phase fluid systems, can exhibit the coalescence of two liquid lenses, as we show schematically in Fig.~(\ref{fig:lens}); recent experiments have shown that, for such a lens merger~\cite{Hack_2020}, $h(t) \sim t^1$ and $h(t) \sim t^{2/3}$ in the viscous and inertial regimes, respectively.
%%%%%%%%%%%%%%%%%%%%%%%%%%%%%%%%%%%%%%%%
%%%%%%%%%%%%%%%%%%%%%%%%%%%%%%%%%%%%%%%%%%
We show that the three-phase Cahn-Hilliard-Navier-Stokes (CHNS3), which couples the fluid velocity $\bm u$ with the concentration fields $c_1, \, c_2,$ and $c_3$, which distinguish between the three coexisting phases that form the lens, provides a natural theoretical framework for the study of liquid-lens coalescence in the viscous and inertial regimes and in the crossover region from the former to the latter. Our direct numerical simulations (DNSs), in both two and three dimensions (2D and 3D), for the coalescence of two nearby, initially static, liquid lenses in this CHNS3 system uncover the 
complete spatiotemporal evolution of $\bm u, \, c_1, \, c_2,$ and $c_3$ during lens mergers. 
In addition, we obtain a variety of new and interesting results that we summarize qualitatively below. We find, in agreement with experiments, that $h(t) \sim t^{\alpha_v}$, with $\alpha_v \simeq 1$ in the viscous regime, which is followed by a region in which the growth of $h(t)$ with $t$ is less steep, and, finally, $h(t) \sim t^{\alpha_i}$, with $\alpha_i \simeq 2/3$ in the inertial regime; we obtain the crossover from the viscous to the inertial regimes as a function of the  Ohnesorge number $Oh$, a dimensionless ratio of viscous stresses to the inertial and surface tension forces~\cite{fardin2022spreading,ohnesorge1936bildung,ohnesorge2019formation} [$Oh \equiv \nu [\rho/(\sigma R_0)]^{1/2}$, where $\rho$, $\nu$, $\sigma$ and $R_0$ are, respectively, the density, viscosity, surface tension, and initial droplet's radius.]. We use the top view of the merger of biconvex lenses in 3D [see the planar section in Fig.~(\ref{fig:lens}) (b)] to define the neck width $w(t)$ and show that  $w(t) \sim t^{\alpha_v}$ and $w(t) \sim t^{\alpha_i}$ in viscous and inertial regions, respectively. From the spatiotemporal evolution of $\bm u, \, c_1, \, c_2,$ and the vorticity $\bm \omega = \nabla \times \bm u$, we demonstrate the crucial role played by a vortex quadrupole that straddles the neck of the merging lenses: the spatial extent of this quadrupole grows with this neck, uniformly in the viscous regime but with distortions in the inertial case, where we see signatures of turbulence, which we quantify by obtaining kinetic-energy and concentration spectra. Such turbulence, during the coalescence of lenses, has not previously been observed in either experimental or numerical studies. We show that the gradient of an excess pressure $P^G_{\mathcal{L}}$ is also of vital importance in the merger of liquid lenses, just as it is in the coalescence of droplets~\cite{fardin2022spreading,liu2022laplace}. Finally, we examine the merger of two asymmetrical, but identical, liquid lenses, whose top parts are more curved than their lower ones. For this asymmetrical case, we exhibit how this proceeds via the coalescence of the upper concave arcs, which is similar to its counterpart for circular droplets, so the growth exponent for $h(t)$ lies in between its lens- and droplet-merger values. To the best of the authors' knowledge, the geometric dependence of such coalescence phenomena has not been previously documented in the scientific literature.   

The remaining part of this paper is organized as follows. In Section~\ref{sec:model}, we define the CHNS3 partial differential equations (PDEs) and the numerical methods we use to solve these PDEs. Section~\ref{sec:results} is devoted to a presentation of our results. We end with concluding remarks in Section~\ref{sec:conclusions}. Section~\ref{sec:appendix} is an Appendix that contains additional figures.
%%%%%%%%%%%%%%%%%%%%%%%%%%%%%%%%%%%%%%%%%%%%%%%%
\begin{figure*}
{\includegraphics[width=\textwidth]{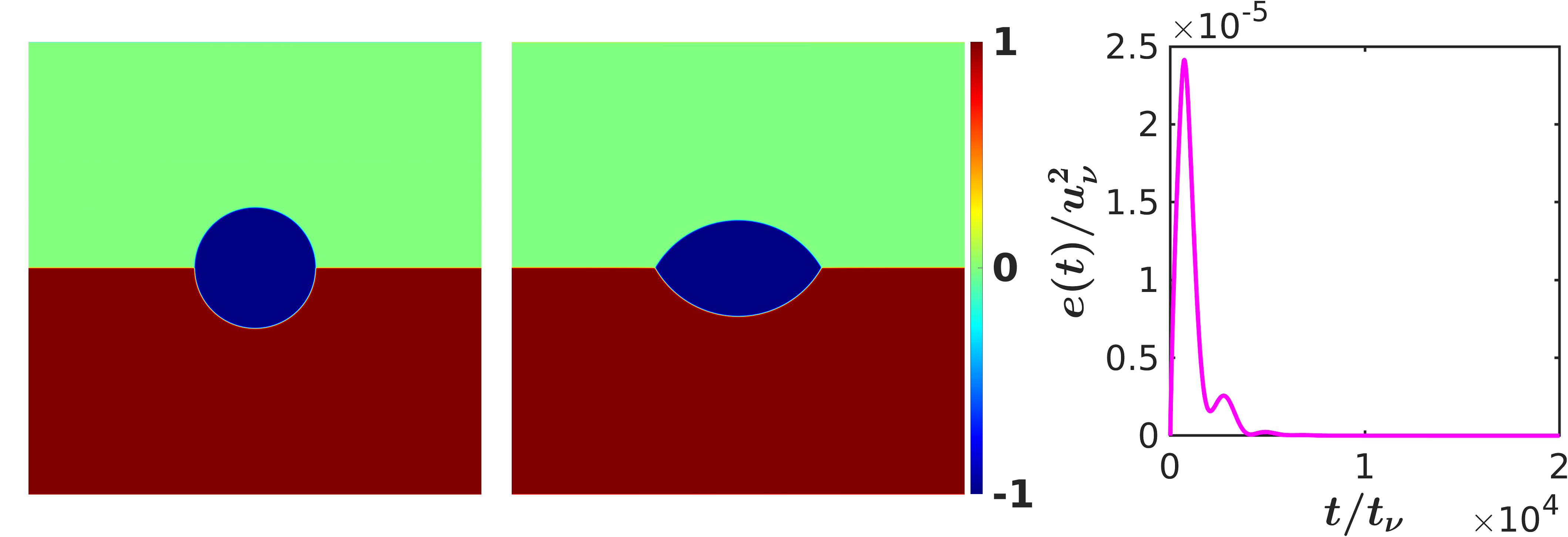} 
\put(-500,150){\rm {\bf(a)}}
\put(-340,150){\rm {\bf(b)}}
\put(-170,150){\rm {\bf(c)}}
}
\caption{\label{fig:init} Pseudocolor plots of $c_2 - c_1$ showing the three co-existing phases and the interfaces between them for (a) the initial condition given in Eq.(~\ref{eq:init}) and (b) the final equilibrium configuration. (c) Plot of the temporal evolution of the kinetic energy $e(t) = \sum_k E(k, t)$ during lens formation for $Oh = 0.09$. The energy is normalized with the viscous scale velocity $u_{\nu} = \sigma/(\rho \nu)$. [We obtain the final equilibrium configuration shown in (b) after the kinetic energy reaches to zero.]}
\end{figure*}
%%%%%%%%%%%%%%%%%%%%%%%%%%%%%%%%%%%%%%%%%%%%%%%
\section{Model and Numerical Methods}
\label{sec:model}
We define the CHNS3 model in Subsection~\ref{subsec:CHNS3}, discuss the details of our direct numerical simulations (DNSs) in Subsection~\ref{subsec:numerics}, and describe the preparation of the lens-merger initial conditions in Subsection~\ref{subsec:IC}.
\subsection{Three-phase Cahn-Hilliard-Navier-Stokes model}
\label{subsec:CHNS3}
Phase-field or Cahn-Hilliard models have been used extensively to study multi-phase fluid flows~\cite{padhan2023activity,roccon2023phase,soligo2020effect,pal2022ephemeral,negro2023yield}; in particular, they have been employed to study droplet coalescence in binary-fluid mixtures~\cite{Gross_2013, Pablo_2010, Pal_2016, palphdthesis}. We show that the following ternary-phase-field (CHNS3) model~\cite{Boyer_2006, Kim_2007, Gyula_2016}, for three immiscible fluids, provides a natural framework for investigations of liquid-lens coalescence; this model uses the variational free-energy functional, in the domain $\Omega$:
%To define three interfaces, we take the following Landau-Ginzburg type free energy functional \cite{Boyer_2006, Kim_2007}:
\begin{eqnarray}
\mathcal{F}(\{c_i, \nabla c_i\}) = \mathlarger \int\displaylimits_{\Omega} d\Omega \left[\frac{12}{\epsilon}F(\{c_i\}) + \frac{3\epsilon}{8} \sum_{i=1}^{3}\gamma_i(\nabla c_{i})^2\right]\,,\;\;\;\;\;
\label{eq:fren}
\end{eqnarray}
where  the concentration fields $c_i (i = 1, 2, 3)$  are conserved order parameters that satisfy the constraint $\displaystyle \sum_{i=1}^{3} c_i = 1$, $\epsilon$ is the thickness of the interface, the variational bulk free energy $F(\{c_i\}) = \displaystyle \sum_{i=1}^{3} \gamma_i c_{i}^2(1-c_{i})^2$, and the gradient terms give the surface-tension penalties for interfaces, with $\sigma_{ij} = (\gamma_i + \gamma_j) / 2$ the bare surface (or interfacial) tension for the interface between the phases $i$ and $j$; the equilibrium values of $c_i$ follow from the global minimum (or minima) of $F(\{c_i\})$. The equilibrium chemical potential of the fluid $i$ is $\mu_i \equiv \delta \mathcal F/\delta c_i + \beta (\{c_i\})$, with $\beta (\{c_i\})$ the Lagrange multiplier that ensures $\displaystyle \sum_{i=1}^3 c_i = 1$, whence we get~\cite{Boyer_2006} 
\begin{eqnarray}
\mu_i &=& -\frac{3}{4}\epsilon \gamma_i \nabla^2 c_i + \frac{12}{\epsilon} [\gamma_i  c_i(1-c_i)(1-2c_i) \nonumber \\
&-& \frac{6\gamma_1 \gamma_2 \gamma_3 (c_1 c_2 c_3)}{\gamma_1 \gamma_2 + \gamma_1 \gamma_3 + \gamma_2 \gamma_3}]\,;
\label{eq:mui}
%\displaystyle \sum_{i=1}^3 \frac{\mu_i}{\gamma_i} &=& 0 \;.
\end{eqnarray}
we \textit{do not} use the summation convention over repeated indices here. The mean fluid velocity $\bm u$ advects the fields $c_i (i = 1, 2, 3)$, which affect the flow, in turn, so that we get~\cite{Boyer_2006} coupled CHNS-type equations for $\bm u$ and $c_1$ and $c_2$ [$c_3$ follows from the constraint $\displaystyle \sum_{i=1}^3 c_i = 1$]. We consider low-Mach-number flows, hence we use incompressible fluids. In 2D it is convenient to use the vorticity-stream-function formulation for the incompressible Navier-Stokes equation to obtain
\begin{eqnarray}
\partial_t{{\omega}} + ({\bm u}\cdot {\nabla}){{ \omega}} &=& \nu {\nabla}^2
{{\omega}} + {\nabla} \times \left({\sum_{i=1}^{3}\mu_{i} {\nabla} c_{i}}\right)\,,\label{eq:2DCHNSA} \\
 \partial_t{c_{j}} + (\bm{u}.{\nabla})c_{j} &=& \frac{M}{\gamma_{j}}{\nabla}^2 \mu_{j}, \;\; j = 1\; \rm{or}\; 2 \,, 
\label{eq:2DCHNSB}
\end{eqnarray}
where we assume, for simplicity, that all the fluids have the same density $\rho = 1$, kinematic viscosity $\nu$, and mobility $M$, and that $\sigma_{12} = \sigma_{23} = \sigma_{13} \equiv \sigma$. In 3D we use
\begin{eqnarray}
\partial_t{\bm u} + ({\bm u}\cdot \nabla){ \bm u} &=& \nu {\nabla}^2
{\bm u} - \nabla P + \left({\sum_{i=1}^{3}\mu_{i} \nabla c_{i}}\right)\,, \label{eq:3DCHNSA}\\
\nabla \cdot \bm u &=& 0\,, \label{eq:3DCHNSB}\\
 \partial_t{c_{j}} + (\bm u.{\nabla})c_{j} &=& \frac{M}{\gamma_{i}}{\nabla}^2 \mu_{j} , \;\; j = 1\; \rm{or}\; 2 \,, 
\label{eq:3DCHNSC}
\end{eqnarray}
where $P$ is the pressure. The terms with $\displaystyle \sum_{i=1}^{3}\mu_{i} \nabla c_{i}$ yield the stress on the fluid because of the fields $c_i$. In addition to the velocity and concentration fields it is instructive to define and evaluate the following:\\
A. The excess pressure $P^G_{\mathcal{L}}$: 
\begin{equation}
\nabla^2 P^G_{\mathcal{L}} = \nabla \cdot \left(\sum_{i=1}^3 \mu_i \nabla c_i\right)\,. \label{equation:Laplace}
\end{equation}
In equilibrium (i.e., no fluid flow) and in the limit of a zero-thickness interface, $P^G_{\mathcal{L}}$ reduces to the conventional Laplace pressure~\cite{Boyer_2006,fisher1984curvature}, which is inversely related to the interface curvature.\\
B. At time $t$, the energy and concentration spectra, the integral scale, and the Reynolds number are, respectively,~\cite{Pal_2016,pandit2017overview,padhan2023activity,Boffetta2012}
\begin{eqnarray}
E(k,t)&=& \frac{1}{2}\displaystyle \sum_{k-1/2<k'<k+1/2} \left[\hat{ \bm{u}}(\mathbf{k}',t)\cdot \hat{ \bm{u}}(-\mathbf{k}',t)\right]\,; \nonumber \\
S_1(k,t) &=& \displaystyle \sum_{k-1/2<k'<k+1/2} |\hat c_1(\mathbf{k}', t)|^2\,,\nonumber \\
S_2(k,t) &=& \displaystyle \sum_{k-1/2<k'<k+1/2} |\hat c_2(\mathbf{k}', t)|^2\,, \nonumber \\
L_I (t) &=& 2\pi \frac{\displaystyle \sum_{k} k^{-1} {E}(k,t)}{\displaystyle \sum_{k} {E}(k,t)}\,, \nonumber \\
Re(t) &=& \frac{U_{rms}(t) L_I(t)}{\nu} \;,
\label{eq:ES1S2LIRe}
\end{eqnarray}
where $U_{rms}(t) = \left[\displaystyle \sum_{k} E(k, t)\right]^{1/2}$ is the root-mean-square velocity of the fluid; $\mathbf{\hat u}(\mathbf{k}',t)$ and $\hat c_i(\mathbf{k}',t)$ are, respectively, the spatial discrete Fourier transforms (DFT) of $\mathbf{u}(\mathbf{x}, t)$ and $c_i(\mathbf{x}, t)$; and $k$ and $k'$ are the moduli of the wave vectors $\mathbf{k}$ and $\mathbf{k}'$.

% %%%%%%%%%%%%%%%%%%%%%%%%%%%%%%%%%%%%%%%%
% %%%%%%%%%%%%%%%%%%%%%%%%%%%%%%%%%%%%%%%%%%%%%%%%
% \begin{figure}
% {\includegraphics[width=9cm]{draft_figures/phi_init.png} 
% \put(-250,110){\rm {\bf(a)}}
% \put(-130,110){\rm {\bf(b)}}
% }
% \caption{\label{fig:init} Pseudocolor plots of $c_2 - c_1$ showing the three co-existing phases and the interfaces between them for (a) the initial condition given in Eq.~\ref{eq:init} and (b) the equilibrium configuration.}
% \end{figure}
% %%%%%%%%%%%%%%%%%%%%%%%%%%%%%%%%%%%%%%%%%%%%%%%
\subsection{Numerical Methods}
\label{subsec:numerics}

We carry out Fourier-pseudospectral DNSs~\cite{Pal_2016,Canuto2012spectral} of the Eqs.~(\ref{eq:2DCHNSA})-(\ref{eq:2DCHNSB}) and Eqs.~(\ref{eq:3DCHNSA})-(\ref{eq:3DCHNSC}) in square ($N^2$ collocation points)  and cubical ($N^3$ collocation points) domains, respectively, with sides $L = 2\pi$, and periodic boundary conditions in all spatial directions. To eliminate aliasing errors, because of the cubic nonlinearity, we use the $1/2$-dealiasing scheme~\cite{Hou_2007} at each time step, before we compute the nonlinear terms in physical space. For time integration, we employ the semi-implicit exponential-time-difference ETDRK2 method~\cite{Cox_2002}. In the CHNS3 model, the fluid velocity and the concentrations $c_i$ change smoothly at fluid interfaces, so we do not have to implement boundary conditions at sharp interfaces. To resolve the interface, we take three grid points in the interface region and we choose $M \simeq \epsilon^2$, so that our phase-field description can approach the sharp-interface limit~\cite{Jacqmin_1999, Yue2010, Magaletti_2013}. The Cahn number $Cn \equiv \epsilon/L$, a non-dimensional measure of the interface width, $R_0/L$, the non-dimensional initial radius of curvature of the lens, and the dimensionless Ohnesorge number $Oh \equiv \nu [\rho/(\sigma R_0)]^{1/2}$ are given in Table~\ref{table:parameters} along with the numbers of collocation points and other parameters for our DNS runs in 2D and 3D. 

Despite the global conservation of the phase-field variable, drops spontaneously undergo shrinking while experiencing shifts from their expected bulk phase values, and these alterations are proportionate to the interfacial thickness~\cite{yue2007spontaneous}. The Cahn numbers we used in all our simulations are very small, for the given computer resolutions; this allows us to preserve the mass conservation of lenses and droplets, to three-decimal-place accuracy. We illustrate area preservation in Fig.~\ref{fig:mass} (see the Appendix) for $Oh = 0.025$ (run 2D-R1), where we plot the ratio $A(t)/A_0$, with $A(t)$ the area of the lenses at time $t$ and $A_0$ their area at the initial time $t = 0$.

\subsection{Initial conditions}
\label{subsec:IC}

To prepare the lens-merger initial condition in 2D for a symmetric and neutrally buoyant lens we start our DNSs with the following configuration for a single circular droplet of fluid 1, with radius $R_0$ and centre $(\pi, \pi)$, placed at the interface between fluids 2 and 3:
\begin{eqnarray}
    c_1(x,y,0) &=& \frac{1}{2}\left[1-\tanh \left(\frac{\sqrt{(x-\pi)^2+(y-\pi)^2}-R_{0}}{2 \sqrt{2} \epsilon}\right)\right];\nonumber\\
    c_2(x, y, 0) &=& \frac{1}{2}\left[1-\tanh \left(\frac{y-\pi}{2 \sqrt{2} \epsilon}\right)\right] - c_1(x,y,0).
    \label{eq:init}
\end{eqnarray}
The initial and equilibrium configurations are similar in 3D.
As time evolves in our DNSs, the initial droplet relaxes to its equilibrium-lens 
(biconvex-lens in 3D) shape as shown in Fig.~\ref{fig:init} for 2D, with the angle $\theta = 120^\circ$ [Fig.~\ref{fig:lens} (a)], because we choose $\sigma_{12} = \sigma_{23} = \sigma_{13} \equiv \sigma$. We then place two such static lenses (biconvex lenses) close 
to each other and set the velocity field to zero everywhere. The initial distance between the proximate edges of the two lenses is greater than the grid spacing $dx$ and less than the interface width $\epsilon$.
%%%%%%%%%%%%%%%%%%%%%%%%%%%%%%%%%%%%%%%%

%%%%%%%%%%%%%%%%%%%%%%%%%%%%%%%%%%%%%
%\nocite{*}
%
%\subsection{Measured quantities}
%

%%%%%%%%%%%%%%%%%%%%%%%%%%%%%%%%%%%%%%%%%%%%%%
\section{Results}
\label{sec:results}
%%%%%%%%%%%%%%%%%%%%%%%%%%%%%%%%%%%%%%%%
\begin{figure*}
 \begin{tikzpicture}
 \node[inner sep=0pt] (image) at (0,0){
 \tcbox[colframe=white!30!black,
            colback=white!30]
{\includegraphics[width = 0.93\textwidth]{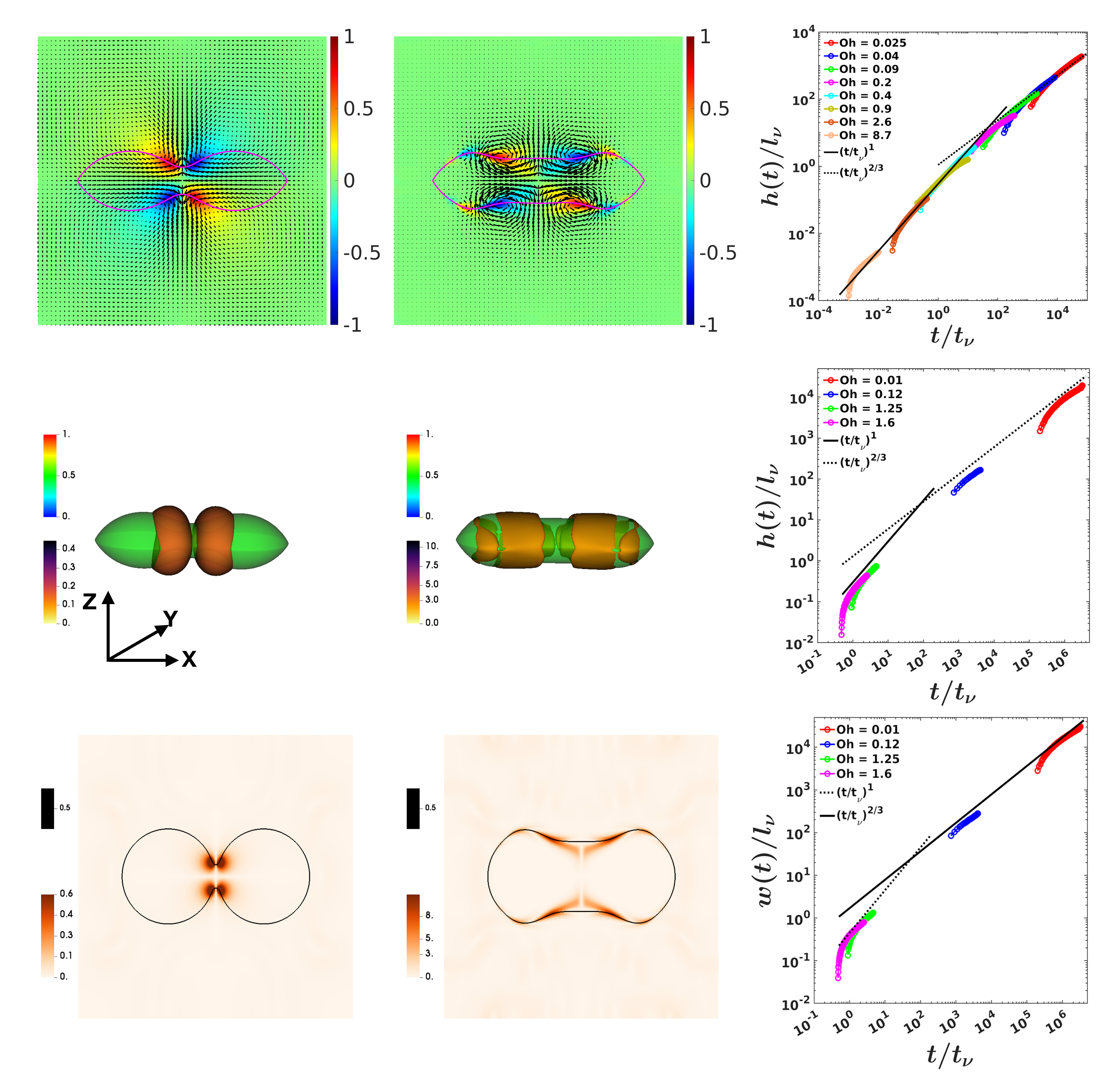}
%\put(-500,510){\rm {$\xrightarrow[\textit{\bf{\large {time}}}]{\hspace*{15cm}}$}}
\put(-485,450){\rm {\bf(a)}}
\put(-485,290){\rm {\bf(d)}}
\put(-485,145){\rm {\bf(g)}}
\put(-320,450){\rm {\bf(b)}}
\put(-320,290){\rm {\bf(e)}}
\put(-320,145){\rm {\bf(h)}}
\put(-150,450){\rm {\bf(c)}}
\put(-150,295){\rm {\bf(f)}}
\put(-150,145){\rm {\bf(i)}}
\put(-470,220){\rm {$\omega$}}
\put(-470,260){\rm {$\phi$}}
\put(-470,60){\rm {$\omega$}}
\put(-470,110){\rm {$\phi$}}
% \put(-360,235){\rm {\bf(j)}}
% \put(-235,235){\rm {\bf(k)}}
% \put(-115,235){\rm {\bf(l)}}
% \put(-480,110){\rm {\bf(m)}}
% \put(-360,110){\rm {\bf(n)}}
% \put(-235,110){\rm {\bf(o)}}
% \put(-115,110){\rm {\bf(p)}}
}};
 \draw[line width=1pt, black] (2.67,-8.32) -- (2.67,8.32);
 \end{tikzpicture}
\caption{\label{fig:lens2D_3D} \textbf{2D DNSs:} Pseudocolor plots of $\bm \omega$ with overlaid velocity vectors for the coalescence of lenses in (a) the viscous regime (multimedia view) [from run 2D-R6] and (b) the inertial regime (multimedia view) [from run 2D-R1]; the $c_1 = 0.5$ contour (magenta line) indicates the lens interface. The field is normalized with its absolute value for ease of visualization. \textbf{3D DNSs:} Isosurface plots of $c_1$ (green) and $|\bm \omega|$ (brown) for (d) the viscous regime (multimedia view) [from run 2D-P3] and (e) the inertial regime (multimedia view) [from run 2D-P1]. \textbf{3D DNSs (top view)} Pseudocolor plots of $\bm \omega(x, y, z=\pi)$ overlaid with the $c_1 = 0.5$ contour line (black line) for (g) the viscous regime [from run 2D-P3] and (h) the inertial regime [from run 2D-P1]. Plots of the scaled neck height $h(t)/l_\nu$ versus the scaled time $t/t_\nu$ for different Ohnesorge numbers 
$Oh$ for (c) 2D lenses (Runs 2D-R1 to 2D-R8, 2D-S1 to 2D-S2) and (f) 3D lenses (Runs 3D-P1 to 3D-P4). (i) Plots of the scaled neck width $w(t)/l_\nu$ versus the scaled time $t/t_\nu$ for different values of 
$Oh$ for the above 3D lenses (top view). The time and length axes are scaled by the corresponding viscous time and length scales. The plots show a clear crossover from the viscous regime, with exponent $\alpha_{v} \simeq 1$, to the inertial regime, with exponent $\alpha_i \simeq 2/3$. In 3D, we measure $h(t)$ and $w(t)$ in the $z$ and $y$ directions, respectively.}
\end{figure*}
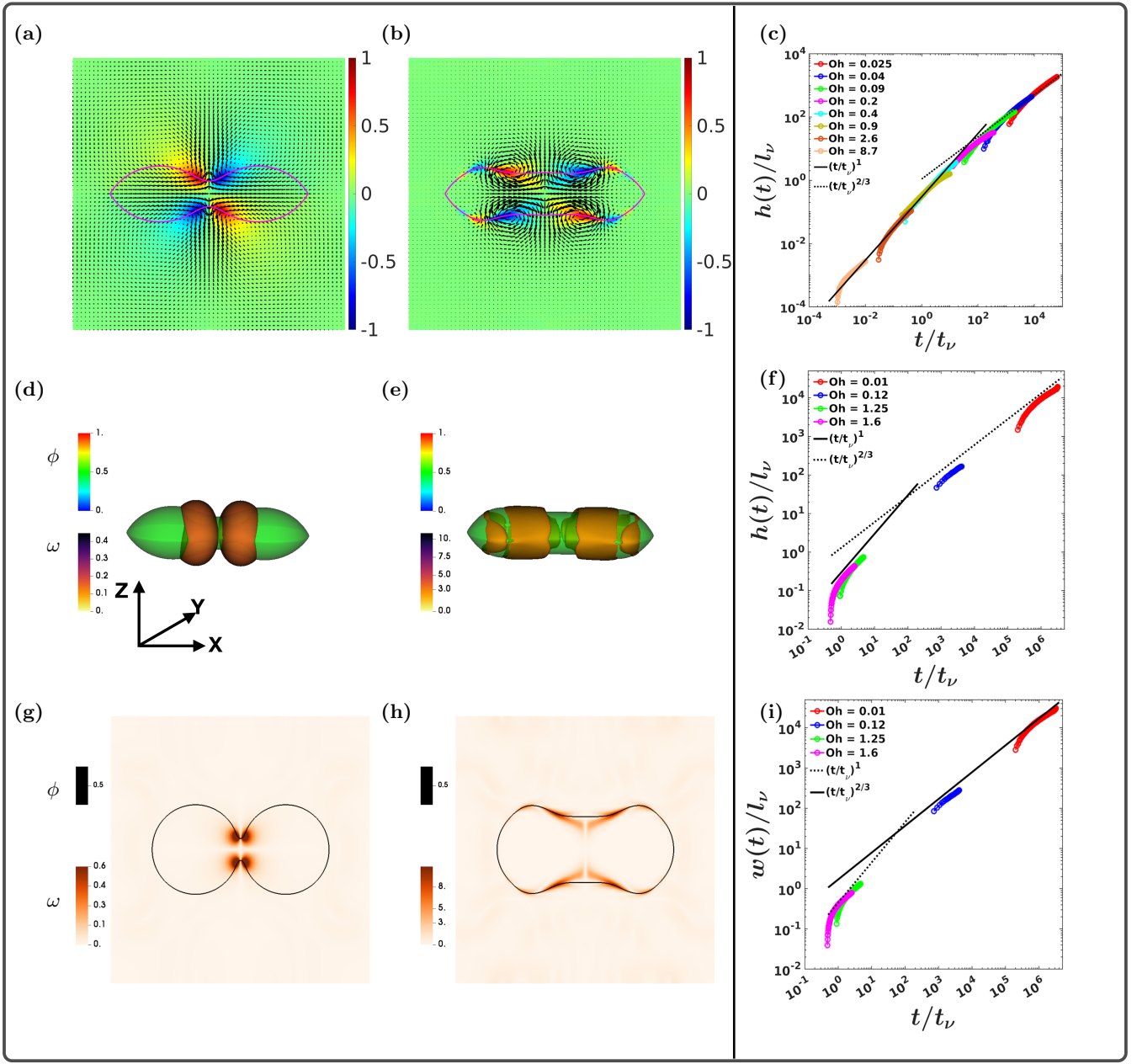
%%%%%%%%%%%%%%%%%%%%%%%%%%%%%%%%%%%%%%%%
%%%%%%%%%%%%%%%%%%%%%%%%%%%%%%%%%%%%%%%%%%%%
\begin{figure*}
{\includegraphics[width=18cm]{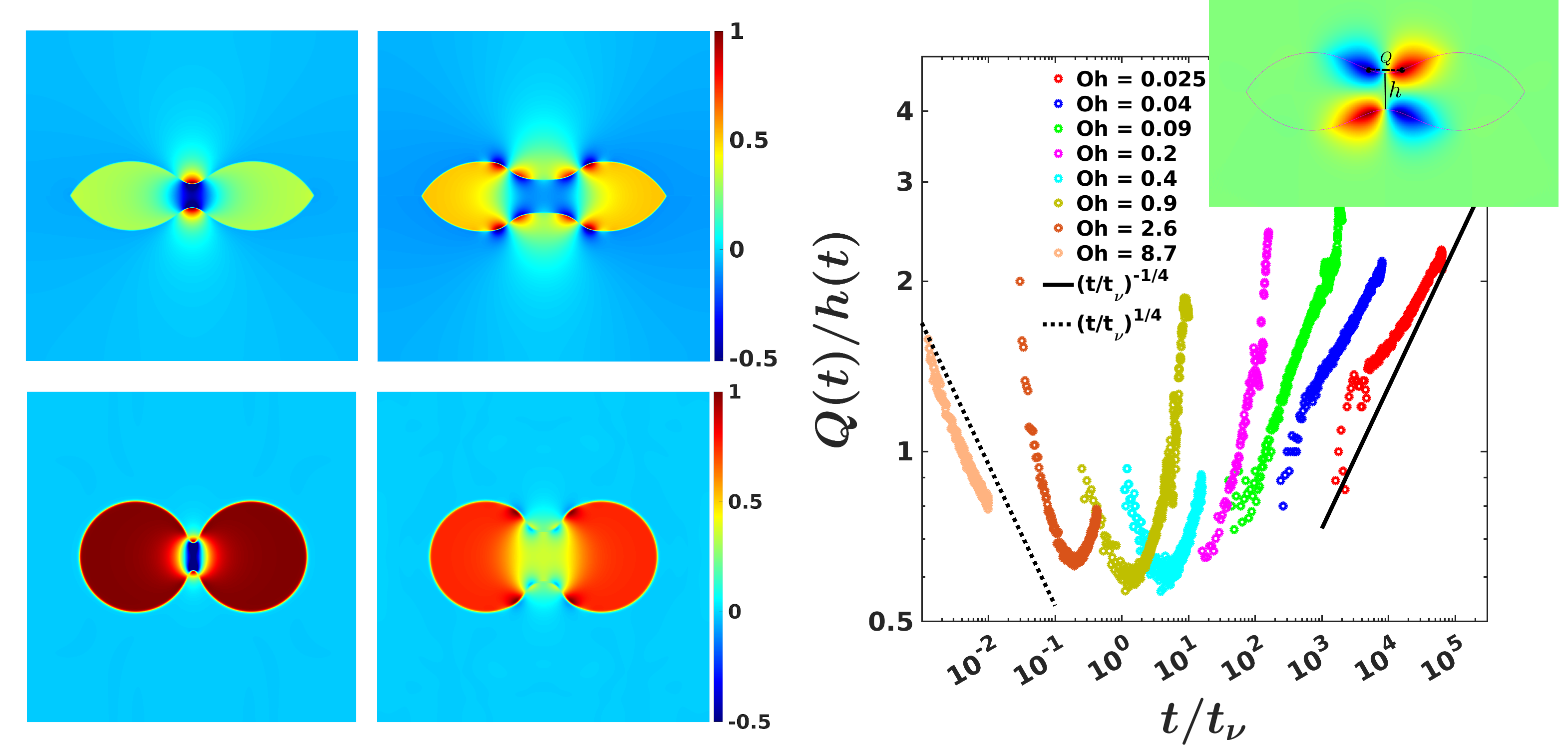}
 \put(-505,225){\rm {\bf(a)}}
 \put(-390,225){\rm {\bf(b)}}
 \put(-505,110){\rm {\bf(c)}}
 \put(-390,110){\rm {\bf(d)}}
 \put(-250,220){\rm {\bf(e)}}
}
\caption{\label{fig:Laplace} Pseudocolor plots of the excess pressure $P^G_{\mathcal{L}}$: For 2D in (a) viscous and (b) inertial regimes. In 3D top view of 
$P^G_{\mathcal{L}}$ in (c) viscous and (d) inertial regimes. (e) Plots versus time $t$ of the ratio of the horizontal width $Q(t)$ of the vortex-quadrupole and the bridge height $h(t)$ (see the top-right schematic figure), for different Ohnesorge numbers $Oh$, showing decay and growth with time (see text) in viscous and inertial regimes.}
\end{figure*}
%%%%%%%%%%%%%%%%%%%%%%%%%%%%%%%%%%%%%%%%%%%%%%%%%
We illustrate the fascinating spatiotemporal evolution of liquid-lens coalescence by representative pseudocolor plots [Figs.~\ref{fig:lens2D_3D} (a) (multimedia view), (b) (multimedia view), (d) (multimedia view), (e) (multimedia view), (g), (h)], from our DNS studies of the symmetrical mergers of two liquid lenses in 2D and of two lenticular biconvex lenses in 3D. In particular, Figs.~\ref{fig:lens2D_3D} (a) (multimedia view) and (b) (multimedia view) show, for the viscous and inertial regimes, respectively, pseudocolor plots of $\bm \omega$, with overlaid velocity vectors and the magenta $c_1 = 0.5$ contour, which is a covenient indicator of the lens interface in 2D. In Figs.~\ref{fig:lens2D_3D} (d) (multimedia view) and (e) (multimedia view), we show results from our DNS of lens mergers in 3D; we use a green isosurface of $c_1$ and an overlaid brown isosurface of $|\bm \omega|$; 
we present $z=\pi$ planar sections of the $c_1$ isosurface (black curve) and of $|\bm \omega|$ (pseuodocolor plots) in Figs.~\ref{fig:lens2D_3D} (g) and (h). We see from these figures and videos that initially static lenses, which are placed close to each other, gradually coalesce by forming a bridge, whose neck height $h(t)$ [and, in 3D, the width $w(t)$ also] increases with the time $t$. This lens coalescence depends on the Ohnesorge number $Oh$. We find, in agreement with experiments~\cite{Hack_2020}, that liquid-lens coalescence is influenced principally by viscous stresses, at high values of $Oh$ (high $\nu$), but by inertial forces, at low values of $Oh$ (low $\nu$), with surface tension forces being the dominant driving factor. We carry out a systematic study of the $Oh$ dependence of this coalescence process.

%%%%%%%%%%%%%%%%%%%%%%%%%%%%%%%%%%%%%%%%%%%%%%%%%
\begin{table}
\caption{The parameters $\nu$, $\sigma$, and $Oh$ for our DNS runs. From 2D-R1 to 2D-R8 and 2D-T1 to 2D-T7  $N^2 = 1024^2$; for runs 2D-S1 to 2D-S2, $N^2 = 2048^2$; for runs 2D-P1 to 2D-P4 $N^3 = 512^3$. For all runs, we use the mobility $M = 10^{-4}$ except for runs 3D-P1 to 3D-P4, for which we take $M = 10^{-3}$. The contact angle $\theta = 120^o$ for all our symmetric-lens simulations. [$*$ In run 2D-K1, with asymmetric lenses, we use $(\sigma_{12}, \sigma_{23}, \sigma_{13}) = (1.4, 1, 0.6); Oh_{ij} = \nu[\rho/(\sigma_{ij}R_0)]^{1/2}$.]}
\label{table:parameters}
\begin{center}
\begin{tabular}{llllll}
\hline
\hline\noalign{\smallskip}
Run ~~~~~~~~~~& $\nu$~~~~~~~~~~~ & $\sigma$~~~~~~~~~ & $R_0/L$~~~~~~~~& $Cn$~~~~~~~~& $Oh$ \\
\noalign{\smallskip}\hline\noalign{\smallskip}
2D-R1 & 0.025 & 1.0 & 0.2 & 0.003& 0.025\\
2D-R2 & 0.05 & 1.0 & 0.2 & 0.003& 0.04\\
2D-R3 & 0.1 & 1.0 & 0.2 & 0.003& 0.09\\
2D-R4 & 0.25 & 1.0 & 0.2 & 0.003& 0.2\\
2D-R5 & 0.5 & 1.0 & 0.2 & 0.003& 0.4\\
2D-R6 & 1.0 & 1.0 & 0.2 & 0.003& 0.9\\
2D-R7 & 5.0 & 1.0 & 0.2 & 0.003& 4.3\\
2D-R8 & 10 & 1.0 & 0.2 & 0.003& 8.7\\

\hline\noalign{\smallskip}

2D-S1 & 3.0 & 1.0 & 0.2 & 0.001& 2.6\\
2D-S2 & 10 & 1.0 & 0.2 & 0.001& 8.7\\

\hline\noalign{\smallskip}

2D-T1 & 0.01 & 1.0 & 0.15 & 0.003& 0.01\\
2D-T2 & 0.05 & 1.0 & 0.15 & 0.003& 0.05\\
2D-T3 & 0.1 & 1.0 & 0.15 & 0.003& 0.1\\
2D-T4 & 0.3 & 1.0 & 0.15 & 0.003& 0.3\\
2D-T5 & 0.5 & 1.0 & 0.15 & 0.003& 0.5\\
2D-T6 & 1.0 & 1.0 & 0.15 & 0.003& 1.0\\
2D-T7 & 2.0 & 1.0 & 0.15 & 0.003& 2.0\\

\hline\noalign{\smallskip}

3D-P1 & 0.01 & 1.0 & 0.1 & 0.006& 0.01\\
3D-P2 & 0.1 & 1.0 & 0.1 & 0.006& 0.12\\
3D-P3 & 1.0 & 1.0 & 0.1 & 0.006& 1.25\\
3D-P4 & 1.25 & 1.0 & 0.1 & 0.006& 1.6\\

\hline\noalign{\smallskip}

2D-K1 & 0.025 & $*$ & 0.2 & 0.003& $*$\\
2D-K2 & 0.025 & 1.0 & 0.2 & 0.003 & 0.025\\

\noalign{\smallskip}\hline
\hline
\end{tabular}
\end{center}
\end{table}
%%%%%%%%%%%%%%%%%%%%%%%%%%%%%%%%%%%%%%%%%%%%%%%%%%%%%%%%%%%%%%
%%%%%%%%%%%%%%%%%%%%%%%%%%%%%%%%%%%%%%%%%%%%%%%%
\begin{figure*}
{\includegraphics[width=18cm]{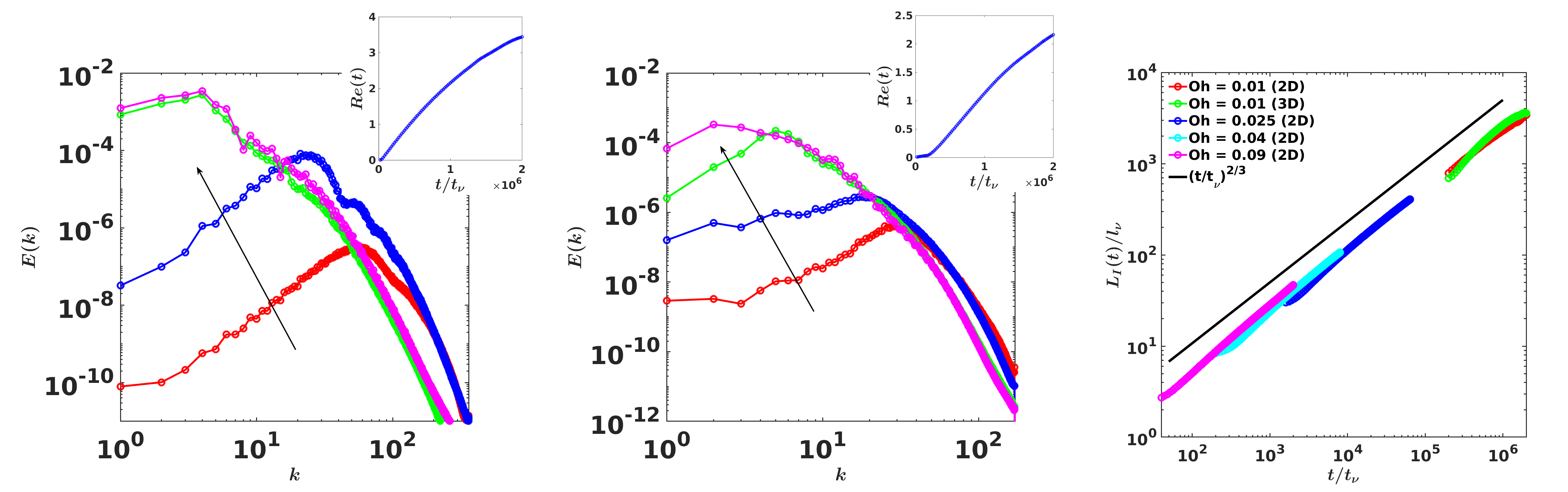}
\put(-510,135){\rm {\bf(a)}}
\put(-340,135){\rm {\bf(b)}}
\put(-160,135){\rm {\bf(c)}}
}
\caption{\label{fig:spectra} Time evolution [at $t/t_{\nu} = 160\; (\equiv t^*) (\text{red line}), \, 10t^* (\text{blue line}), \, 50t^* (\text{green line}),\, {\rm{and}} \, 300t^* (\text{magenta line})$] of the inertial-regime-kinetic-energy spectra $E(k,t)$ for (a) run 2D-T1 in 2D and (b) run 3D-P1 in 3D; the insets on the top right show the Reynolds number $Re(t)$. (c) In the inertial regime, plots of $L_I(t)/l_\nu$ versus $t/t_\nu$ collapse significantly and indicate
power-law scaling with $L_I(t)/l_\nu \sim [t/t_\nu]^{\alpha_L}$ with the scaling exponent $\alpha_L \simeq 2/3$.}
\end{figure*}
%%%%%%%%%%%%%%%%%%%%%%%%%%%%%%%%%%%%%%%%%%%%%%%
In Figs.~\ref{fig:lens2D_3D} (c), (f), (i), we quantify the remarkable difference between the growth of $h(t)$ in the viscous and inertial regimes. In both 2D and 3D, our DNSs yield $h(t)/l_{\nu} \sim (t/t_{\nu})^{\alpha_v}$ and $h(t)/l_{\nu} \sim (t/t_{\nu})^{\alpha_i}$ with distinctly different viscous- and inertial-range exponents $\alpha_v\simeq 1$ and $\alpha_i\simeq 2/3$, respectively; here, $l_{\nu} = \rho \nu^2/\sigma$ and $t_{\nu} = \rho^2\nu^3/\sigma^2$ are the viscous length and time scales\cite{Dirk_2005}. Our results are in consonance with recent experiments on the coalescence of liquid lenses~\cite{Hack_2020}. Figures~\ref{fig:lens2D_3D} (c) and (f) demonstrate clearly that, if we plot the scaled neck height $h(t)/l_{\nu}$ versus the scaled time $t/t_{\nu}$, then the curves for different values of $Oh$ collapse, to a significant degree, onto a single curve, whose asymptotes are the viscous- and inertial-range scaling forms mentioned above; these asymptotes are separated by a broad crossover region. Within the accuracy of our measurements (and those in experiments) the scaling exponents $\alpha_v$ and $\alpha_i$ are universal insofar as they \textit{do not depend on} $Oh$ and the linear size and the spatial dimension of the symmetrical lenses [see Fig.~\ref{fig:size} in the Appendix]. Furthermore, as we show in Fig.~\ref{fig:lens2D_3D} (i), in 3D the scaled width $w(t)/l_{\nu}$ also shows the collapse, for different values of $Oh$, and the same scaling forms as $h(t)/l_{\nu}$.

 We find that, in viscous-regime coalescence, neck growth is guided by the large gradient of $P^G_{\mathcal{L}}$ [see Figs.~\ref{fig:Laplace}(a) and (c) for 2D and the top view for 3D, respectively]. In contrast, in the inertial regime, 
 %the neck-growth exponent is $2/3$ because 
 the gradient of $P^G_{\mathcal{L}}$ [Eq.~\ref{equation:Laplace}], in the region of the neck, is smaller than it is in the viscous case  [see Figs.~\ref{fig:Laplace}(b) and (d) for 2D and the top view for 3D, respectively]. This leads to faster neck growth in the viscous case than in the inertial one, with $\alpha_v \simeq 1 > \alpha_i \simeq 2/3$.
 %an exponent of $1$. This indicates that the high-pressure gradient at the neck region results %in a more rapid merging of the fluid lenses. In contrast, in the inertial regime, the neck-%growth exponent is $2/3$ because the gradient of $P^G_{\mathcal{L}}$ in this region is smaller %than it is in the viscous case  [see Fig.~\ref{fig:Laplace}(b) for 2D and (d) for the top view %for 3D]. 

The following heuristic dimensional argument~\cite{Eddi_2013} suggests why the exponents $\alpha_v$ and $\alpha_i$ are different from each other. On dimensional grounds, $\nabla P^G_{\mathcal{L}}  \sim P^G_{\mathcal{L}}/h(t)$. The velocity of growth of the neck height is ${\dot h(t)}$. In the viscous regime $\nu \nabla^2 \bm{u} \sim \nu \dot h(t)/ h^2$; if we balance this by $\nabla P^G_{\mathcal{L}} \sim P^G_{\mathcal{L}}/h(t)$ and note that $P^G_{\mathcal{L}} \sim \sigma/h$, we obtain $\nu \dot h(t) \sim \sigma$, whence $h(t)\sim t$ and $\alpha_v = 1$. If we equate the inertial term $\bm{u} \cdot \bm{\nabla} \bm{u}$
with $\dot{h}^2/h$, the balance with $\nabla P^G_{\mathcal{L}} \sim P^G_{\mathcal{L}}/h(t)$ yields $h(t) \sim t^{2/3}$, i.e., $\alpha_i = 2/3$. The exponent $\alpha_i=2/3$, for inertial-range liquid-lens coalescence, is distinct from its counterpart in the coalescence of spherical droplets, where $\alpha_i=1/2$ [see, e.g., Refs.~\cite{Thomas_2004, Dirk_2005, Burton_2007, Paulsen_2011, Eggers_1999, Eggers_2003, Volker_2018, Gross_2013, Akella_2020, palphdthesis}]. This indicates that the geometry of the coalescing droplets plays a major role in the coalescence process, as has been noted in recent experiments~\cite{Hack_2020}.

The superimposition of the vorticity and velocity fields, which we present in Figs.~\ref{fig:lens2D_3D}(a),(b),(d),(e), (g), and (h), shows clearly that, in the viscous regime, a vortex quadrupole
is present in the region of the neck of the vortex. In the inertial case, this quadrupole stretches out with some subsidiary small vortices; the neck of the lens stretches out also in this case. In the inertial case, the presence of numerous vortices, spread over the interface, is indicative of turbulence~\cite{Boffetta2012,pandit2017overview}, whose properties we explore below. 

We investigate the spreading of the vortex quadrupole and its distortion into a pair of dipoles by computing the ratio $Q(t)/h(t)$, where $Q(t)$ is the distance between the vortex and anti-vortex cores [see the top inset of Fig.\ref{fig:Laplace}(e)]. In the log-log plots of Fig.~\ref{fig:Laplace} (e), we show how $Q(t)/h(t)$ varies with time $t$ for different values of $Oh$. In the viscous regime, $Q(t)/h(t)$ decreases as $t$ increases; by contrast, in the inertial regime, $Q(t)/h(t)$ increases with $t$. At the highest (lowest) value of $Oh$ that we consider, this decrease (increase) is characterized by a power-law exponent $\simeq -1/4$ ($\simeq +1/4$); for intermediate values of $Oh$, the ratio $Q(t)/h(t)$ first decreases and then increases as $t$ progresses. 

%The neck grows both vertically and horizontally in 3D coalescence, which is distinct from its %2D counterpart, where there exists only vertical growth of the neck. Similar to the vertical %neck growth, the growth in the horizontal directions $w(t)$ also shows scaling exponents $1$ %and $2/3$ in the viscous and inertial regimes, as shown in Fig.\ref{fig:lens2D_3D}(i). The %spatiotemporal evolution of the vorticity field illustrated in Figs.\ref{fig:lens2D_3D}(c),(f) %is consistent with Fig.~\ref{fig:lens2D_3D}(a),(b). 

%%%%%%%%%%%%%%%%%%%%%%%%%%%%%%%%%%%%%%%%%%%%
\begin{figure}
\hspace{-1cm}
\includegraphics[width=9.5cm]{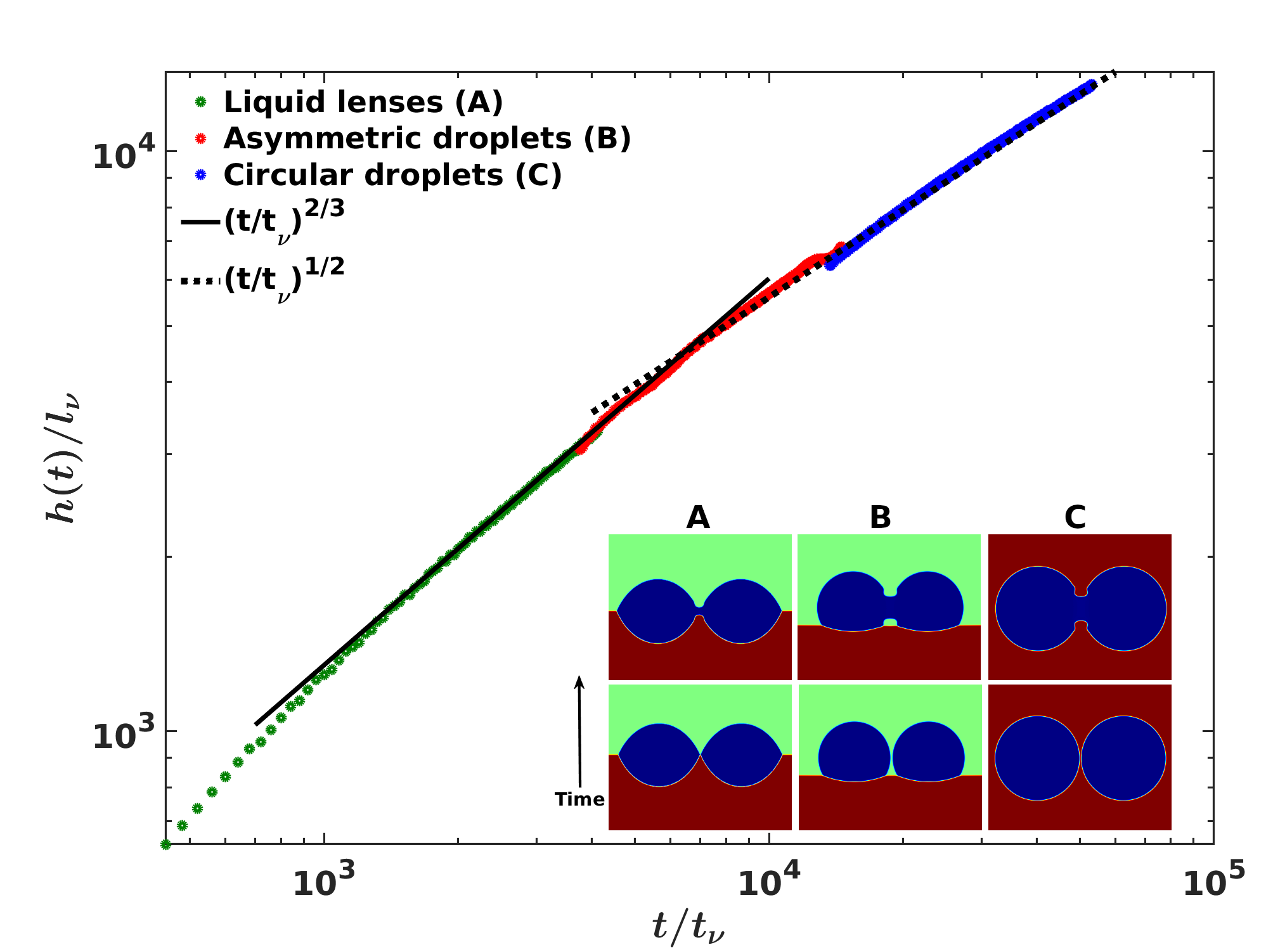}
\caption{\label{fig: geometry} Pseudocolor plots of $c_2-c_1$ (see insets) illustrating the mergers of (A) symmetric lenses (multimedia view), (B) asymmetric lenses (multimedia view), and (C) circular droplets (multimedia view) (in a two-phase system with, say, $c_1=0$). Log-log plots of $h(t)/l_\nu$ versus $t/t_\nu$ for the mergers in (A), (B), and (C) illustrating the geometry dependence the power-law-growth exponent $\alpha_i$ in the inertial regime. For the coalescence of symmetric lenses [from run 2D-R1] $\alpha_i \simeq 2/3$ (green line); this value shows a smooth crossover to $\alpha_i \simeq 1/2$ (blue line) for the coalescence of circular droplets [from run 2D-K2]; the coalescence of asymmetric lenses [from run 2D-K1] shows a crossover from $\alpha_i \simeq 2/3$ to $\alpha_i \simeq 1/2$ (red line).}
\end{figure}
%%%%%%%%%%%%%%%%%%%%%%%%%%%%%%%%%%%%%%%%%%%

Significant turbulence is generated during liquid-lens coalescence in the inertial regime. We quantify this turbulence by considering the temporal evolution of the energy spectrum $E(k,t)$, which yields the energy distribution across different wave numbers $k$, the integral length scale $L_I(t)$, which is the typical length scale of energy-containing eddies, and the Reynolds number $Re(t)$ that characterizes the degree of turbulence [see Eq.~\ref{eq:ES1S2LIRe}].

 In Fig.~\ref{fig:spectra} (a) [2D run T1] and Fig.~\ref{fig:spectra} (b) [3D run P1] we present log-log plots of $E(k,t)$ versus $k$ for several representative times $t$; the embedded figures on the top right corners show the growth of $Re(t)$ with $t$. From these figures we see that the energy is spread over at least two decades over $k$; this is a clear signature of lens-merger-induced turbulence. The arrows in Fig.~\ref{fig:spectra}(a) and (b) indicate the direction of time evolution of the energy spectra during coalescence, suggesting inverse cascades of energy in both 2D and 3D.[The concentration spectra $S_1(k,t)$ and $S_2(k,t)$ are also spread over at least two decades of $k$ because of this turbulence (see Fig.~\ref{fig:Sk} in the Appendix), but their dependence on $t$ is less than that of $E(k,t)$.] The time evolution of the scaled integral length scale $L_I(t)/l_\nu$, shown in Fig.~\ref{fig:spectra} (c), indicates power-law scaling with $L_I(t)/l_\nu \sim [t/t_\nu]^{\alpha_L}$, with the scaling exponent $\alpha_L \simeq 2/3$  [this is like the neck-growth exponent shown in Figs.~\ref{fig:lens2D_3D}(c), (f), and (i)]. 

In Fig.~\ref{fig: geometry} we compare log-log plots of $h(t)/l_\nu$ versus $t/t_\nu$ for the mergers of (A) symmetric lenses [see Fig.~\ref{fig: geometry}A (multimedia view)], (B) asymmetric lenses [Fig.~\ref{fig: geometry}B (multimedia view)], and (C) circular droplets [Fig.~\ref{fig: geometry}C (multimedia view)]. [See the pseudocolor plots of $c_2-c_1$ in the insets. Symmetric lenses, as illustrated in Fig. 6(A), exhibit top-down symmetry; in contrast, asymmetric lenses, as depicted in Fig. 6(B), do not display this symmetry.] These plots demonstrate the geometry dependence of the power-law-growth exponent $\alpha_i$ in the inertial regime. Specifically, we find: for the coalescence of symmetric lenses  [run 2D-R1] $\alpha_i \simeq 2/3$ (green line); this value shows a smooth crossover to $\alpha_i \simeq 1/2$ (blue line) for the coalescence of circular droplets [from run 2D-K2]; the coalescence of asymmetric lenses [from run 2D-K1] shows a crossover from $\alpha_i \simeq 2/3$ to $\alpha_i \simeq 1/2$ (red line).

\section{Conclusion and Discussions} 
\label{sec:conclusions}

We have shown that the three-phase Cahn-Hilliard-Navier-Stokes (CHNS3) provides a natural theoretical framework for the study of liquid-lens coalescence in the viscous and inertial regimes and in the crossover region from the former to the latter. By carrying out extensive DNSs, we have shown, in agreement with experiments, that (a) $h(t) \sim t^{\alpha_v}$, with $\alpha_v \simeq 1$ in the viscous regime; (b) in the crossover region the growth of $h(t)$ with $t$ is less steep; and (c) $h(t) \sim t^{\alpha_i}$, with $\alpha_i \simeq 2/3$ in the inertial regime. Our study of the viscous, crossover, and inertial regimes as a function of $Oh$ and $R_0$ have demonstrated that these exponents are universal and do not depend on the sizes of the merging lenses. From the top view of the merger of biconvex lenses in 3D [Fig.~(\ref{fig:lens}) (b)] we have shown that  $w(t) \sim t^{\alpha_v}$ and $w(t) \sim t^{\alpha_i}$ in viscous and inertial regions, respectively. By monitoring the spatiotemporal evolution of $\bm u, \, c_1, \, c_2,$ and $\bm \omega$, we have uncovered the crucial role played by a vortex quadrupole in this merger; and we have characterized the growth and distortion of this 
quadrupole. In the inertial case, we have unveiled signatures of lens-merger-induced turbulence, which we have quantified via the spectra $E(k,t),\, S_1(k,t),$ and $S_2(k,t)$, and $L_I(t)$ and $Re(t)$. We have shown that the gradient of $P^G_{\mathcal{L}}$ is of importance in lens mergers, just as it is in the coalescence of droplets~\cite{fardin2022spreading,liu2022laplace}. 
Our examination of the merger of two asymmetrical lenses has elucidated how this proceeds via the coalescence of the upper concave arcs, so the growth exponent $\alpha_i$ lies in between its lens- and droplet-merger values. We hope that our detailed study of the spatiotemporal evolution of concentration and velocity fields during liquid-lens mergers will lead to experimental investigations of this evolution and of lens-merger-induced turbulence.

%Our DNSs of the coalescence of two liquid lenses by using the CHNS3 formalism. Our studies revealed that when two lenses coalesce, a neck grows with time and whose growth follows self-similar behaviours. In the viscous regime such growth follows a power-law with an exponent $1$ associated with a vortex quadrupole at the neck region leading to high pressure gradient at the neck. On the contrary, the inertial coalescence is dominated by the breaking and spreading of the vortex quadrupole leading to relatively low pressure gradient at the neck, and it follows a power-law with an exponent $2/3$. We showed that such difference in the scaling exponents is related to the interfacial spreading of the vortices, which we quantified by calculating the ratio between the vortex distance and the neck height. The ratio deceases when the flow is in the viscous regime and increases when it is in the inertial regime. 

%\section{Phase field spectra}

 We note that liquid-lens coalescence is often studied for sessile droplets on solid substrates in many experiments~\cite{Thoroddsen_2005, Lee_2012, Eddi_2013, Sui_2013}. It is possible to study the spatiotemporal evolution of such coalescence by combining our CHNS framework with a volume-penalization scheme as we will show elsewhere.

 As we were preparing our paper for publication, we became aware of another paper, which has just been published recently ~\cite{scheel2023viscous}, that has carried out a Lattice-Boltzmann study of symmetric liquid-lens mergers in 2D and 3D. This study obtains results that are similar to those that are summarised in our Fig.~\ref{fig:lens2D_3D}.

\section*{Acknowledgments}
We thank Jaya Kumar Alageshan and Nairita Pal for valuable discussions and the Science and Engineering Research Board (SERB) and the National Supercomputing Mission (NSM) Grant No. $\text{DST/NSM/R\&D\_
HPC\_Applications/2021/34}$, India for support, and the Supercomputer Education and Research Centre (IISc) for computational resources.  

\section*{Data and code availability}

Data from this study and the computer scripts can be obtained from the authors upon 
reasonable request.

\section*{Conflicts of Interest}
No conflicts of interests, financial or otherwise, are declared by the authors.

\section*{Author Contributions} NBP and RP planned the research and analysed the numerical data; NBP carried out the calculations and prepared the tables, figures, and the draft of the manuscript; NBP and RP then revised the manuscript in detail and approved the final version.

\begin{widetext}
\section{Appendix}
\label{sec:appendix}
%%%%%%%%%%%%%%%%%%%%%%%%%%
% \begin{figure*}[ht!]
%     \centering
%     \begin{minipage}{0.45\linewidth}
%     \includegraphics[width=\linewidth]{figure_7.png}
% \caption{The neck growth for lenses of initial radius of curvature $R_0/L = 0.15$. The power-law exponents $\alpha_v$ and $\alpha_i$ for different $Oh$ [from runs 2D-T1 to 2D-T7] are independent of the size of the lenses.}
% \label{fig:size}
%     \end{minipage}
%     \begin{minipage}{0.6\linewidth}
%     \includegraphics[width=\linewidth]{area_lens_en.png}
% \caption{The plot showing the ratio $A(t)/A_0$ versus time, where $A(t)$ is the area of the lenses at time $t$ and $A_0$ is the area of the same lenses at the initial time $t = 0$.}
% \label{mass}
%     \end{minipage}
% \end{figure*}
%%%%%%%%%%%%%%%%%%%%%%%%%%
\begin{figure}
{\includegraphics[width=8.5cm]{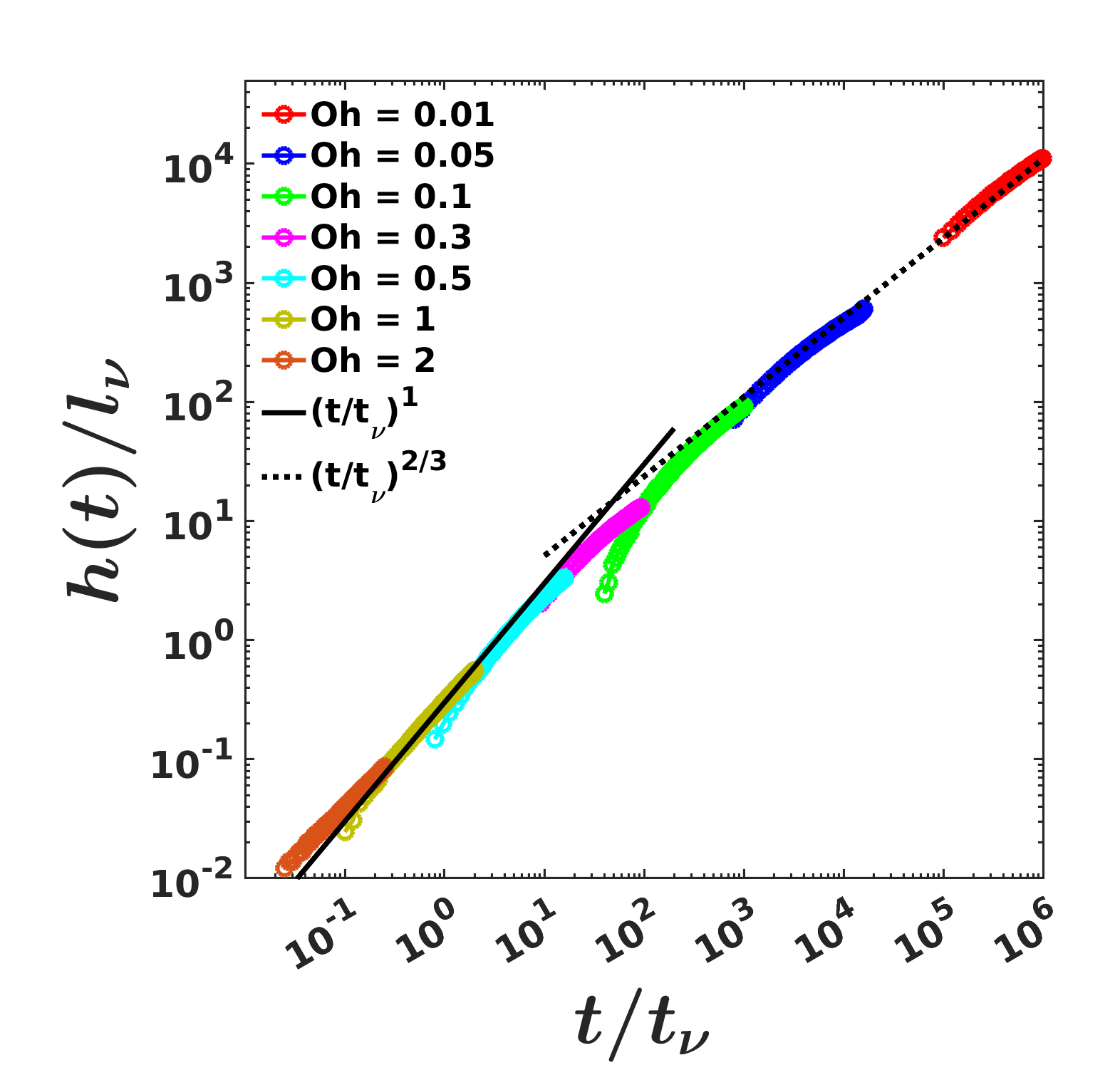}
}
\caption{\label{fig:size} The neck growth for lenses of initial radius of curvature $R_0/L = 0.15$. The power-law exponents $\alpha_v$ and $\alpha_i$ for different $Oh$ [from runs 2D-T1 to 2D-T7] are independent of the size of the lenses.}
\end{figure}
%%%%%%%%%%%%%%%%%%%%%%%%%%%%%%%%%%%%%
 The concentration spectra $S_1(k,t)$ and $S_2(k,t)$ are also spread over at least two decades of $k$ because of lens-merger-induced turbulence [see Fig.~\ref{fig:Sk}], but their dependence on $t$ is less than that of $E(k,t)$.
%%%%%%%%%%%%%%%%%%%%%%%%%%%%%%%%%%%%%
\begin{figure*}
{\includegraphics[width=18cm]{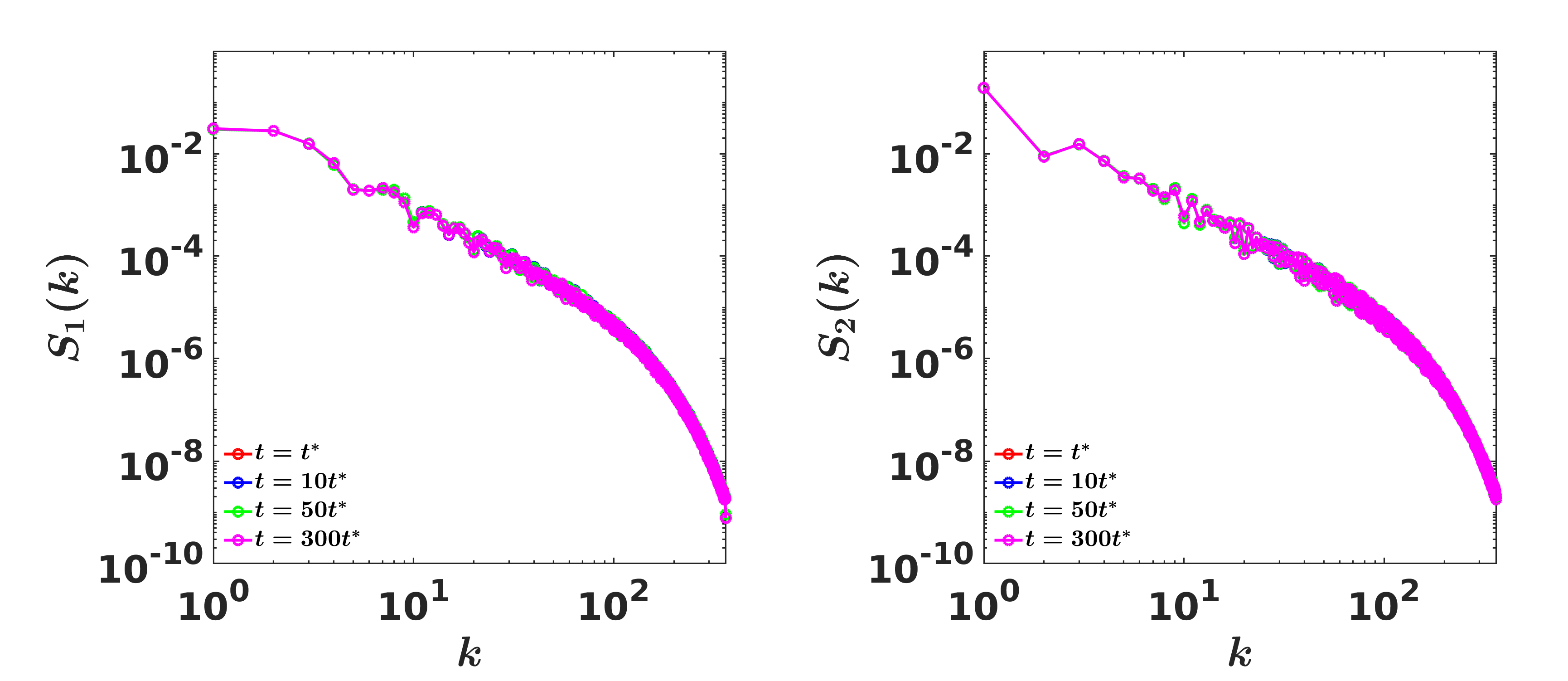}
\put(-480,200){\rm {\bf(a)}}
\put(-220,200){\rm {\bf(b)}}
}
\caption{\label{fig:Sk} Time evolution [at $t/t_{\nu} = 160 (\equiv t^*) (\text{red line}),\, 10t^* (\text{blue line}),\, 50t^* (\text{green line}),\, {\rm{and}}\, 300t^* (\text{magenta line})$] of the phase-field spectra (a) $S_1(k,t)$ and (b) $S_2(k, t)$ [from run 2D-T1].}
\end{figure*}
%%%%%%%%%%%%%%%%%%%%%%%%%%%%%%%%%%%%%
%%%%%%%%%%%%%%%%%%%%%%%%%%%%%%%%%%%%%%%%%%%%%%%%
\begin{figure}
{\includegraphics[width=0.5\textwidth]{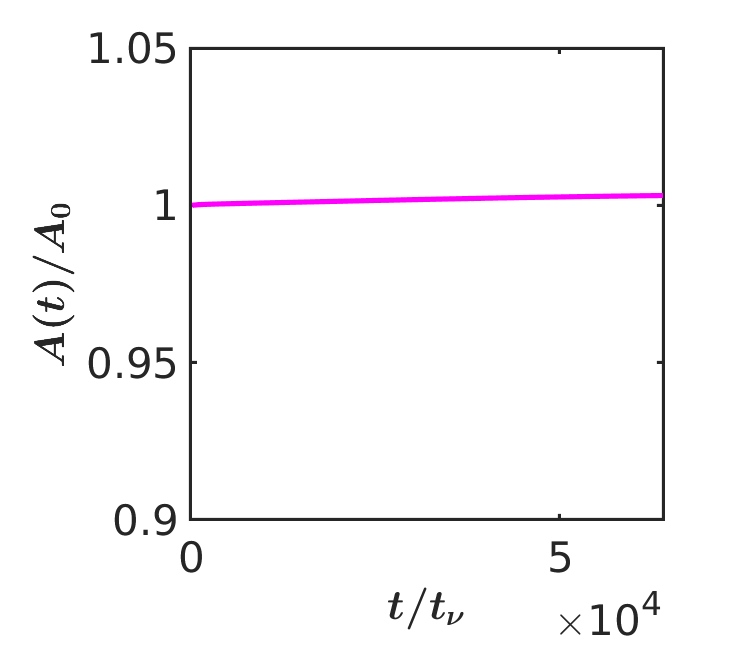} 
}
\caption{\label{fig:mass} The plot showing the ratio $A(t)/A_0$ versus time, where $A(t)$ is the area of the lenses at time $t$ and $A_0$ is the area of the same lenses at the initial time $t = 0$.}
\end{figure}
%%%%%%%%%%%%%%%%%%%%%%%%%%%%%%%%%%%%%%%%%%%%%%%
\end{widetext}

%\nocite{*}

%\bibliography{lens}% Produces the bibliography via BibTeX.
\bibliographystyle{apsrev4-2}

%\begin{bibliography}
\providecommand{\noopsort}[1]{}\providecommand{\singleletter}[1]{#1}%
%
%\end{bibliography}
\end{document}